\documentclass[a4paper, 11pt]{article}

\usepackage[american]{babel}
\usepackage{csquotes}
\usepackage[style=apa, citestyle=numeric, backend=biber, sorting=none,
	maxbibnames=99, giveninits=true]{biblatex}

\DeclareFieldFormat{labelnumberwidth}{\mkbibbrackets{#1}}

\defbibenvironment{bibliography}
  {\list
     {\printtext[labelnumberwidth]{%
      \printfield{labelprefix}%
      \printfield{labelnumber}}}
     {\setlength{\labelwidth}{\labelnumberwidth}%
      \setlength{\leftmargin}{\labelwidth}%
      \setlength{\labelsep}{\biblabelsep}%
      \addtolength{\leftmargin}{\labelsep}%
      \setlength{\itemsep}{\bibitemsep}%
      \setlength{\parsep}{\bibparsep}}%
      }
  {\endlist}
  {\item}
  
\usepackage[utf8]{inputenc}
\usepackage{algorithm,algpseudocode}
\usepackage{appendix}
\usepackage{authblk}
\usepackage{booktabs}			
\usepackage{enumitem}   

\usepackage{fancyvrb}			
\usepackage{float}			
\usepackage[T1]{fontenc}		
\usepackage{hyperref}			
\usepackage{manyfoot}
\usepackage{microtype}			
\usepackage[normalem]{ulem}		
\usepackage{url}			
\usepackage{amsfonts,amsmath,amssymb,amsthm}
\usepackage{bm,bbm}			
\usepackage{mathrsfs}			
\usepackage{mathtools}			
\usepackage[super]{nth}			
\usepackage{oubraces}			
\usepackage{stmaryrd}			

\usepackage{graphicx}
\usepackage[caption=false]{subfig}
\usepackage{tikz}
\usepackage{tkz-graph}
\usetikzlibrary{cd, matrix}
\usetikzlibrary{decorations.pathreplacing}

\theoremstyle{plain}%
\newtheorem{theorem}{Theorem}
\newtheorem{lemma}[theorem]{Lemma}%
\newtheorem{proposition}[theorem]{Proposition}%

\theoremstyle{remark}%
\newtheorem{example}{Example}%
\newtheorem{remark}{Remark}%

\theoremstyle{definition}%
\newtheorem{definition}{Definition}%

\begin{document}

\title{New Construction of Locally q-ary Sequential Recoverable Codes: Parity-check Matrix Approach}

\author[1]{Akram Baghban}
\author[2]{Mehdi Ghiyasvand}
\affil[1,2]{Department of Mathematics, Faculty of Science, Bu-Ali Sina University \\ \text{a.baghban@sci.basu.ac.ir, mghiyasvand@basu.ac.ir}}

\maketitle{}

\begin{abstract}
This paper develops a new family of locally recoverable codes for distributed storage systems, Sequential Locally Recoverable Codes (SLRCs) constructed to handle multiple erasures in a sequential recovery approach.
We propose a new connection between parallel and sequential recovery, which leads to a general construction of q-ary linear codes with information $(r, t_i, \delta)$-sequential-locality where each of the $i$-th information symbols is contained in $t_i$ punctured subcodes with length $(r+\delta-1)$ and minimum distance $\delta$. 
We prove that such codes are $(r, t)_q$-SLRC ($t \geq \delta t_i+1$), which implies that they permit sequential recovery for up to $t$ erasures each one by $r$ other code symbols.

 	\section*{Keywords} Distributed storage systems, Local recovery, Parallel and Sequential recovery.
\end{abstract}


\section{Introduction}

The distributed storage system (DSS) employs various types of coding methods to prevent the problems of straightforward replication of data.
Among them, the locally recoverable codes (LRCs) \cite{bib1}, have received a lot of interest in recent times. The main aim of using them is to minimize the total number of nodes needed for node recovery by utilizing nearby encoded components instead of the entire dataset.
A $q$-ary LRC, called $[n, k]_q$-LRC, with locality $r$ is a linear code over a finite field $\mathbb{F}_q$, where the recovery of each code symbol is achievable by accessing at most $r$ other symbols.
The \textit{sequential recovery} and the \textit{parallel recovery} are the two approaches that are capable of dealing with the issues of local recovery for failures of multiple nodes in DSS systems (\cite{bib10}, \cite{bib11}). 

Failed nodes are only repairable from the remaining unfailed nodes when applying the parallel approach \cite{bib4}. Some well-known classes of them include codes with $(r, t)$-locality, in which each code symbol has $t$ pairwise disjoint repair sets with locality $r$ (\cite{bib12}, \cite{bib16}, \cite{kazemi2023upper}); Codes with $(r, \delta)$-locality, in which each information symbol and/or each code symbol is contained in a local code of length at most $r+\delta-1$ and minimum distance at least $\delta$ (\cite{bib7}, \cite{bib14}); and codes with overall local repair erasures $t$, in which for any $\mathcal{I} \subseteq [n]$ of size at most $t$ and any $i \in \mathcal{I}$, the $i$-th code symbol has a repair set with locality $r$ contained in $[n]\backslash E$ \cite{bib4}, were proposed. 
For $(r, t)$-locality codes that have locality $r$ and availability $t$, for all code symbols, the well-known code rate
bound is presented in \cite{bib12}:
\begin{equation}\label{eq:1}
    \frac{k}{n}\leq \prod^{t}_{j=1}\frac{1}{1+\frac{1}{jr}}.
\end{equation}
In the sequential approach, $t$ erasures ($t \geq 2$) can be ordered in a sequence and one--by--one recovered. Consequently, the previously repaired nodes can be employed for repairing the remaining failed nodes (\cite{bib11}, \cite{bib2}, \cite{bib13}). A code with $(r, t)$-sequential-local-recovery, called $(r, t)$-SLRC, is defined as an $[n, k]$ linear code where any $t$-seq erasures can be recovered, each one by $r$ other code symbols. 
The sequential recovery has a much greater advantage in terms of erasure tolerance than the parallel recovery with the same $n$ and $k$ \cite{bib17}.

In recent studies, several methods have been utilized to propose constructions of $(r, t)$-SLRCs for both binary and/or non-binary $t$-seq LRCs.
Such as graph (\cite{bib11}, \cite{bib13}), resolvable configuration \cite{bib8}, code products (\cite{bib8}, \cite{bib20}), good polynomial \cite{bib22} and designing the parity-check matrix (\cite{bib11}, \cite{bib8}, \cite{bib13}).
The binary codes have less complexity and are simple to implement (\cite{bib11}, \cite{bib13}, \cite{bib8}), whereas the non-binary codes provide a larger minimum distance and a greater number of recoverable nodes (\cite{bib20}, \cite{bib22}). 

The first two constructions (i.e. graph and configuration) are rate optimal for binary $2$-seq and $3$-seq LRCs.
In \cite{bib11} and \cite{bib8}, the rate-optimal for $(r, t=2)$-SLRCs is provided as:
\begin{equation}\label{eq:2}
    \frac{k}{n} \leq \frac{r}{r+2}.
\end{equation}
For codes with $(r, t=3)$-SLRCs, the upper bound on the code rate satisfies (\cite{bib8}, \cite{bib2}, \cite{bib9}):
\begin{equation}\label{eq:3}
    \frac{k}{n} \leq (\frac{r}{r+1})^{2}.
\end{equation}
Researchers in \cite{bib13} introduced bounds on the code rate and the block length of $t$-seq LRCs. They derived the parity-check matrix structure for a rate-optimal LRC with sequential recovery, enabling a graphical code representation. Furthermore, they identified a family of binary codes associated with particular regular graphs, known as \textit{Moore graphs}.
The authors of \cite{bib8} created a code construction based on resolvable configurations using the parity-check matrix method, which results in the limitation odd $t \geq 3$ for binary $t$-seq LRCs. 
The code rate of this family satisfies:
\begin{equation}\label{eq:4}
    \frac{k}{n}=(1+\frac{t-1}{r}+\frac{1}{r^2})^{-1}.
\end{equation}
On the other hand, in the case of non-binary codes for $t$-seq LRCs, the authors of \cite{bib20} used the direct product approach to construct a new and interesting group of $q$-ary linear $(r, t)$-SLRCs, which can recover any number of erasures by a small locality $r$.
In addition, they improved the upper bound on the total number of recoverable nodes in comparison with earlier studies.
The other construction of non-binary codes with even locality $r$ for handling $2$-seq LRCs was approached in study \cite{bib22}, which utilizes \textit{good polynomials} over a small alphabet size $q$, potentially resulting in rate optimality.
Moreover, paper \cite{bib21} established a connection between sequential recovery and linear block codes.
It demonstrated that a linear block code with girth $2(t+1)$ could function as a $t$-seq LRC with locality $r$ if its parity-check matrix had columns weighing at least $2$ and rows weighing at most $r+1$.

Despite all this, it is still evident that the non-binary issue for $(r, t)$-SLRCs receives little attention.
Based on considering the parity-check matrix technique and inspired by research \cite{bib16} and \cite{bib8}, we introduce a construction for $t$-seq LRCs with locality $r$, establishing a new connection between parallel and sequential recovery. This leads to q-ary linear codes with information $(r, t_i, \delta)$-sequential-locality, which are $(r, t)_q$-SLRCs.
Significantly, present construction, reduces ambiguity and complexity, in the binary $(r, t)$-SLRCs presented in \cite{bib13}.
In addition, it eliminates the conditions on locality $r \geq 3$ and even mentions them in \cite{bib13} and \cite{bib22}.
Moreover, it removes the requirement for $2$-seq and $t$ to be odd, as shown in (\cite{bib8}, \cite{bib22}).
\cite{kazemi2023upper}


\section{Preliminaries and Notation}

For any set $A$, the notation $|A|$ denotes the number of elements of $A$. 
If $B$ is a subset of $A$ with $|B| = r$, then $B$ is referred to as a $r$-subset of $A$.
For any positive integer $n$, let $[n]$ represent the
$\{1, 2, ..., n\}$.
In the context of a q-ary linear code $\mathcal{C}$, denoted as an $[n, k]_q$, and $\mathcal{I} \subseteq [n]$, $\mathcal{C}|_{\mathcal{I}}$ denotes the punctured subcode obtained by removing coordinates in $\overline{\mathcal{I}} := [n]\backslash \mathcal{I}$.
The support $supp(x)$ of a vector $x = \left(x_1, x_2, ..., x_n\right) \in \mathcal{C}$ is the set of coordinates of its non-zero entries, and its weight is defined to be $wt(x)= |supp(x)|$.
Let $i \in [n]$ and $ \mathcal{R}_i \subseteq [n] \setminus \{i\}$.
The subset $\mathcal{R}_i$ is called an $r$-recovery set of $i$ if $|{\mathcal{R}_i}| \leq r$ and for all $(x_1, ..., x_n) \in \mathcal{C}$, $ x_i=\sum_{j\in \mathcal{R}_i}{a_{ij}x_{j}}$, where all $a_{ij}\in \mathbb{F}_q$ and are independent of the value of $x_i$.
Moreover, in a linear $[n, k]_q$ code $\mathcal{C}$, a code symbol is said to have $(r, t')$-availability if there exist $t'$ disjoint groups of other symbols, each with size at most $r$, to recover this symbol.

\begin{definition}\label{def:1}
    The $i$-th code symbol $c_i$, in a linear code $\mathcal{C}$, is said to have $(r, \delta)$-locality if there exists a punctured subcode $\widehat{\mathcal{C}}_i$ of $\mathcal{C}$ with support containing $i$, which has length at most $r+\delta-1$ and minimum distance at least $\delta$. Equivalently, there exists a subset $\mathcal{R}_i\subseteq [n]$ such that 
     \begin{itemize}
        \item $i\in \mathcal{R}_i$, and $|\mathcal{R}_i|\leq r+\delta-1$,
        \item The minimum distance of the subcode $\mathcal{C}|_{\mathcal{R}_i}$ obtained by puncturing code symbols in $\overline{\mathcal{R}_i}:=[n]\backslash \mathcal{R}_i$ equals $\delta$.
    \end{itemize}
\end{definition}

\begin{definition}\label{def:2}\cite{bib20}
    A $q$-ary linear $[n,k]_q$ code $\mathcal{C}$ is called $(r, t)_q$-Sequential-Locally-Recoverable Code (SLRC) over a finite field $\mathbb{F}_q$ if, for each $\mathcal{I} \subseteq [n]$ of size at most $t$, the elements of $\mathcal{I}$ can be arranged sequentially as $\{i_1, i_2, ..., i_t \}$. 
    Moreover, for each $i_j \in \mathcal{I}$, exists a recovery set $ \mathcal{R}_j \subseteq \overline{\mathcal{I}}\cup\{i_1, i_2, ..., i_{j-1}\}$, of size at most $r$, where $\overline{\mathcal{I}} := [n] \backslash \mathcal{I}$.  
\end{definition}

\begin{proposition}\label{pro:1}
\cite{bib8} Code $\mathcal{C}$ is an $(r, t)_q$-SLRC if and only if for any nonempty $\mathcal{I} \subseteq [n]$
of size $|\mathcal{I}|\leq t$, there exists an $i \in \mathcal{I}$ such that $i$ has a recovery set $\mathcal{R}_i \subseteq [n] \backslash \mathcal{I}$.
\end{proposition}

The lemma below generalizes a similar structure for the case of $\ell = 2$, presented in \cite{bib8}.

\begin{lemma}\label{lem:1}
    Suppose for $1 \leq j \leq \ell$, there exists nonempty $\ell$ pairwise disjoint sets $A_1, ..., A_{\ell}$,
    such that $[n] = \bigcup_{j=1}^{\ell}A_j$ and for each $j$, $|A_j|\leq t_{j}$ where $t_{j} \geq 0$.
    Then if $\mathcal{C}$ is a q-ary linear $[n, k]_q$ code such that
\begin{enumerate}
    \item  For any nonempty $\mathcal{I} \subseteq A_j$ of size $|\mathcal{I}| \leq t_{j}$, there exists an $i \in \mathcal{I}$ such that $i$ has a recovering set $\mathcal{R}_i \subseteq A_j \backslash \mathcal{I}$;
    \item  For any nonempty $\mathcal{I} \subseteq A_{j}$ of size $|\mathcal{I}| \leq t_1 + ...+ t_{j}+1$, there exists an $i \in \mathcal{I}$ such that $i$ has a recovering set $\mathcal{R}_i \subseteq [n]\backslash \mathcal{I}$;
\end{enumerate} 
    Then, $\mathcal{C}$ is an $(r, t)_q$-SLRC with $t=1+\sum_{j=1}^{\ell}t_j$.
\end{lemma}


\section{Information $(r, t_i, \delta)$-Sequential-Locally-Recoverable Code}

In this section, we will discuss a new q-ary family of linear sequential locally recoverable codes that has $(r, t_i, \delta)$-sequential-locality for each of its information symbols, which is a novel concept in the sequential recovery approach and a new connection between parallel recovery and sequential recovery codes.
Following \cite{bib16} and \cite{bib8}, we combine the notion of $(r, \delta)$-locality with $(r, t)_q$-SLRCs, considering a situation where the number of parity symbols in each local subcode is fixed and equal to the upper bound on the number of erasures that can be recovered.

In other words, each of $i$-th information symbol of these codes, for $1 \leq i \leq k$, is included in $t_i$ punctured subcodes such that each of these subcodes is a $(r, \delta)_q$-LRC and permits parallel recovery for up to $\delta$ erasures every one by $r$ other code symbols.
Accordingly, we will establish that our general code construction is a non-binary $(r, t)_q$-SLRC capable of recovering up to $t$ failed nodes, satisfying the condition $t \geq \delta t_i+1$. 
This implies that we are primarily interested in focusing on a situation, where we do not have disjoint recovery sets in the global code and sequential recovery performs better than parallel recovery.

We employ the subsequent definition to provide the above details as the main concept of this research.

\begin{definition}\label{def:3}
    A linear $[n, k]_q$ code $\mathcal{C}$ is called an information $(r, t_i, \delta)$-sequential-locally-recoverable code (SLRC) if each information code symbol of $\mathcal{C}$ is contained in $t_i$ punctured subcodes $\widehat{\mathcal{C}_1}$, ..., $\widehat{\mathcal{C}}_{t_i}$,   each with length at most $r+\delta-1$ and minimum distance $\delta$, and their supports only intersect on this coordinate. 
    Additionally, each local subcode is an $(r, \delta)_q$-locality code and contains exactly $\delta-1$ parity symbols such that $t_i \leq \delta$.
    
    Equivalently, for each $i \in [k]$, there exist $t_i$ subset $\mathcal{R}_1, ..., \mathcal{R}_{t_i} \subseteq [n]$ such that 
 
 \begin{enumerate}
   \item  $i\in \mathcal{R}_j$, and $|\mathcal{R}_j|\leq r+\delta-1$, $\forall j \in [t_i]$;
   \item The minimum distance of the subcode $\mathcal{C}|_{\mathcal{R}_j}$ obtained by puncturing code symbols in $\overline{\mathcal{R}_j}:=[n]\backslash \mathcal{R}_j$ equals $\delta$, $\forall j \in [t_i]$;
   \item $\mathcal{R}_j \cap \mathcal{R}_{j'}=\{i\}$, for any $j \neq j' \in [t_i]$;
  \item $|\mathcal{R}_j \cap \{k+1, k+2, ..., n\}|= \delta-1$, for each $j\in [t_i]$;
  \item for any $\mathcal{I} \subseteq [n]$ with $|\mathcal{I}| \leq \delta t_i +1$, there exists an $i \in \mathcal{I}$ such that $i$ has a recovery set $\mathcal{R}_i \subseteq [n]\backslash \mathcal{I}$.
\end{enumerate}

\end{definition}

An important point for the reader to consider in the above definition is that $t_i$ is dependent on $i$ and could change in $i$.
To better understand the $q$-ary $(r, t_i, \delta)$-SLRC construction, the following example is presented.

\begin{example}\label{ex:1}
 Suppose $r=3$, $t_i=2$, and $\delta=3$. Additionally, let $\mathcal{C}$  represent a linear $[16, 6]_q$ code over the finite field $\mathbb{F}_4$ with the parity-check matrix $H$ as defined below.
  
\begin{equation}\label{eq:5}
{H= \left( \begin{array}{cccccc|cccccccc|cc} 
            1 & 1 & 1 & 0 & 0 & 0 & 1 & 0 & 0 & 0 & 0 & 0 & 0 & 0 & 0 & 0 \\
            1 &\beta &\beta^2 & 0 & 0 & 0 & 0 & 1 & 0 & 0 & 0 & 0 & 0 & 0 & 0 & 0 \\
            
            1 & 0 & 0 & 1 & 1 & 0 & 0 & 0 & 1 & 0 & 0 & 0 & 0 & 0 & 0 & 0\\
            1 & 0 & 0 &\beta &\beta^2 & 0 & 0 & 0 & 0 & 1 & 0 & 0 & 0 & 0 & 0 & 0 \\
            
            0 & 1 & 0 & 1 & 0 & 1 & 0 & 0 & 0 & 0 & 1 & 0 & 0 & 0 & 0 & 0 \\
            0 & 1 & 0 &\beta & 0 &\beta^2 & 0 & 0 & 0 & 0 & 0 & 1 & 0 & 0 & 0 & 0 \\
            0 & 0 & 1 & 0 & 1 & 1 & 0 & 0 & 0 & 0 & 0 & 0 & 1 & 0 & 0 & 0 \\
            0 & 0 & 1 & 0 &\beta &\beta^2 & 0 & 0 & 0 & 0 & 0 & 0 & 0 & 1 & 0 & 0\\
            \hline
            0 & 0 & 0 & 0 & 0 & 0 & 1 & 1 & 1 & 0 & 0 & 0 & 0 & 0 & 1 & 0\\
            0 & 0 & 0 & 0 & 0 & 0 & 1 & \beta & \beta^2 & 0 & 0 & 0 & 0 & 0 & 0 & 1 
    \end{array} \right)}
 \end{equation}
By the first $8$ rows of $H$, it can be observed that each information symbol (coordinate) is contained in two local subcodes that only intersect at this information symbol.
For instance, when considering the initial information symbol, we define the following local sets: 
$\mathcal{R}_1=\{1, 2, 3, 7, 8\}$ and 
$\mathcal{R}_2=\{1, 4, 5, 9, 10\}$.
Both sets contain $5$ elements, satisfying the condition $|\mathcal{R}_j|\leq r+\delta-1$ (i.e., $5\leq 5$). Furthermore, these local subcodes must maintain a minimum distance of $\delta$.

For the last part of the definition, we assume that the recovery sets associated with the local subcodes are as follows: The recovery sets for the first coordinate can include $\{2, 3, 7\}$, $\{2, 3, 8\}$, $\{4, 5, 9\}$, and $\{4, 5, 10\}$;
For the coordinates related to recovery, we also have $\mathcal{R}_7=\{1, 2, 3\} \subseteq \{1,...,6 \}$, $\mathcal{R'}_{7}=\{8, 9, 15\}$, and for the coordinate $15$, there exists a recovery set $\mathcal{R}_{15}=\{7, 8, 9\}$.
Each of these sets must have size $r=3$ and be a subset of $[n]\backslash \mathcal{I}$, where $\mathcal{I}$ denotes the erased symbols.
Therefore, the linear $[16, 6]_4$ code $\mathcal{C}$ is a $(3, 2, 3)$-SLR code.
\end{example}

\begin{remark}
Based on what is proposed as an LRC in \cite{bib16} (Lemma 1), the upper bound of the failed information nodes that can be repaired is at most $\delta t_i-1$ node failures, while by considering the sequential recovery approach, we can recover up to $\delta t_i+1$ node failures.
\end{remark}
In this paper, we are mainly interested in the case, where we have disjoint recovery sets for the information code symbols and do not have them for other code symbols, and where sequential recovery is beneficial over parallel recovery.
In Section 4, we explain the process of obtaining the above parity check matrix $H$. For this purpose, we need to discuss some essential concepts.

\begin{definition}\label{def:4}
    A binary matrix $M$ is considered $(m, n)$-regular if it has uniform column weight $m$ and row weight $n$. Furthermore, if the supports of any two rows of $M$ intersect on at most one common coordinate, $M$ is said to have girth \textbf{g}$\geq 4$. 
\end{definition}

\begin{remark}
    There exist diverse structures of $(m, n)$-regular matrices with girth \textbf{g}$\geq 4$. One important group of such matrices are the incidence matrices of certain objects from block designs, e.g., the $2$ design.
\end{remark} 
Moved by this idea, we recall the definition of \textit{resolvable configurations} from \cite{bib8}, which is a $2$ design.

\begin{definition}\label{def:5}\cite{bib8}.
    Let $\mathcal{P}$ be a set of $k$ elements, called  Points, and $\mathcal{L}$ be a collection of subsets of $\mathcal{P}$, called lines. The pair of $(\mathcal{P},\mathcal{L})$ is known as a $(t_i,r)$-Resolvable-Design if the following conditions are met:
  \begin{itemize}
     \item Each line contains $r$ points;
     \item Each point belongs to $t_i$ lines;
     \item Every pair of distinct lines has at most one point in common;
     \item All lines in $\mathcal{L}$ can be partitioned into $t_i$ parallel classes, where a parallel class is a set of lines that partition $\mathcal{P}$.
   \end{itemize}
\end{definition}
In all cases $(t_i, r)$-resolvable-design, each corresponding parallel class contains $\lceil\frac{k}{r}\rceil$ lines and $b=\lceil\frac{k}{r} \rceil t_i$.
The incidence matrix of a $(t_i, r)$-resolvable-design in which, 
$\mathcal{P}=\{p_1, ..., p_{k}\}$ and $\mathcal{L}=\{L_1, ..., L_{b}\}$, 
is described as a $b \times k$ binary matrix $M=(m_{i,j})$ where
\begin{equation}
m_{i,j} = \left\{ 
\begin{array}{rcl}
           1, & \mbox{if}\ x_j \in B_i \\
           0, & \mbox{otherwise}.
\end{array}\right.    
\end{equation} 
Obviously, $M$ is a $(t_i, r)$-regular matrix with girth \textbf{g}$\geq 4$.

\begin{example}\label{ex:2}
Let $k=6$ and $\mathcal{P}=\{p_1, p_2, ..., p_6\}$. 
Let $r=3$ and $t_i=2$.
Also, $\mathcal{L}=\{L_1, L_2, L_3, L_4\}$, where $L_{1}=\{p_1, p_2, p_3\}$, $L_{2}=\{p_1, p_4, p_5\}$, $L_{3}=\{p_2, p_4, p_6\}$ and $L_{4}=\{p_3, p_5, p_6\}$. 
Then, $\mathcal{A}_1=\{L_1, L_2\}$ and $\mathcal{A}_2=\{L_3, L_4\}$ are two disjoint classes of $(\mathcal{P},\mathcal{L})$, 
and every pair of lines from disjoint classes intersects at just one point.
Therefore, we obtain its incidence matrix presented as a following binary $(2, 3)$-regular matrix with \textbf{g}$\geq 4$:
\begin{equation}\label{eq:6}
     M_{4 \times 6}=
     \left( \begin{array}{cccccc} 
            1 & 1 & 1 & 0 & 0 & 0\\
            1 & 0 & 0 & 1 & 1 & 0\\
            0 & 1 & 0 & 1 & 0 & 1\\
            0 & 0 & 1 & 0 & 1 & 1
            \end{array} \right).
 \end{equation}
\end{example}
Algorithm 1 develops our information $(r, t_i, \delta)$-sequential-locality code construction.

\section{Algorithm 1}

In this section, we were inspired by the LRC construction described in \cite{bib16} and established a new family of q-ary sequential recovery codes that can recover up to $t$ failed nodes and satisfy the definition \ref{def:3}.

\noindent\rule{\textwidth}{0.4pt}
\textbf{Algorithm 1:} Process of constructing the parity-check matrix $H$ of the information $(r, t_i, \delta)$-sequential-locality code

\noindent\rule{\textwidth}{0.4pt}

\begin{description}
    \item [\textbf{Start:}]
    \item [\hspace{0.4cm}\textbf{Inputs:}]
    \item \hspace{0.5cm}\textbf{Step 1:}
        Take $\mathbb{F}_q$ as a finite field such that $q \geq r+\delta-2$ and $\beta$, as a primitive element.
        Take $I_{m}$, the identity matrix, and $O_{m'}$, the all-zero matrix for any positive integers $m=\ell \times \ell$ and $m'=\ell \times \ell'$.
        
    \item \hspace{0.4cm} \textbf{Step 2:} 
        Input $\widehat{H}=[Q_{(\delta-1)\times r}, I_{(\delta-1)}]$
        as a full-rank parity-check matrix of a q-ary $[r+\delta-1, r, \delta]$-MDS code over $\mathbb{F}_q$.

    \item \hspace{0.4cm} \textbf{Step 3:} 
        Input $M$ as an incidence matrix of $(\mathcal{P},\mathcal{L})$, which is a $b \times k$ binary $(t_i, r)$-regular matrix with uniform column weight $t_i$, uniform row weight $r$, and girth \textbf{g}$\geq 4$.

   \item [\hspace{0.4cm}\textbf{Procedure:}]
   \item \hspace{0.5cm}\textbf{Step 1:} 
   \textit{Derive Matrix $M^*:$}
         For $j=1\ \text{to}\ r$, follow these steps:
        \begin{itemize}
         \item Go through every row of matrix $M$.
         \item Replace the $j$-th 1 in that row with the $j$-th column vector of $Q$ in $\hat{H}$.
         \item Replace any $0$'s with column vectors filled with zeros of size $(\delta - 1)$.
         \item Output: $b(\delta-1)\times k$.
        \end{itemize}
    
    \item \hspace{0.5cm}\textbf{Step 2:} 
      \textit{Derive Matrix $W^*:$}
       \begin{itemize}
           \item Calculate $s=\lceil\frac{k}{r}\rceil$ from the parameters of $M$.
           \item Partition $[s]$ into $\lceil\frac{s}{r}\rceil$ nonempty subsets.
           \item Derive $W^*$ such that
                \begin{equation}\label{eq:7}
                 W^*=\left( \begin{array}{ccc}
                   Q^1 & & O_{m}\\
                       & \ddots &  \\
                       O_{m'}&  & Q^{\lceil\frac{s}{r}\rceil}
                  \end{array} \right).
                \end{equation}
           \item Output: $\lceil\frac{s}{r}\rceil(\delta-1)\times \lceil\frac{s}{r}\rceil r$.      
        \end{itemize}
    \item \hspace{0.5cm}\textbf{Step 3:} 
        \textit{Derive Matrix $H:$}
          \begin{itemize}
            \item Combining $M^*$ and $W^*$ together such as below
                \begin{equation}\label{eq:8}
                    H=\left( \begin{array}{c|c|c}
                    M^* & I_{b(\delta-1)} & O_{m}\\
                        \hline
                    O_{m'} & W^* \ \  O_{m''} & I_{\lceil \frac{s}{r}\rceil(\delta-1)}\\
                             \end{array} \right).
               \end{equation}
            \item Put $m=b(\delta-1) \times \lceil\frac{s}{r}\rceil(\delta-1)$.
            \item Put $m'=\lceil\frac{s}{r}\rceil(\delta-1) \times k$.
            \item Put $m''=\lceil\frac{s}{r}\rceil(\delta-1) \times b(\delta-1) - \lceil\frac{s}{r}\rceil r$.
        \end{itemize}
    \item [\hspace{0.4cm}\textbf{Output:}] \textit{Return parity-check matrix $H$}.
    \item [\textbf{End.}]
         \noindent\rule{\textwidth}{0.4pt}
     \end{description}

\begin{remark}
Clearly, $H$ has $(b+\lceil\frac{s}{r}\rceil)(\delta-1)$ rows and $n=k+(b+\lceil\frac{s}{r}\rceil)(\delta-1)$ columns. So the rank of $H$ is $b(\delta-1)$. 
\end{remark}

\begin{example}\label{ex:3}
Let $q=4$ and $\beta$ be primitive elements of $\mathbb{F}_4$.
Suppose $r=3,\ \delta=3$, and
 \begin{equation}\label{eq:9}
     \widehat{H}=\left( \begin{array}{ccc|cc} 1 & 1 & 1 & 1 & 0 \\
                                1 & \beta & \beta^2 & 0 & 1
            \end{array} \right)
 \end{equation}
is a parity-check matrix of a $[5, 3, 3]$-MDS code linked to \textit{Step 2} of Inputs in Algorithm $1$.
Additionally, consider the incidence $(2, 3)$-regular matrix in Example \ref{ex:2}, corresponding to \textit{Step 3} of that Algorithm. 
Consequently, we can derive $M^*_{8\times6}$ through the procedure outlined in \textit{Step 1} of the Algorithm's Procedure section, as described below
\begin{equation}\label{eq:10}
     M^*_{8 \times 6}=
     \left( \begin{array}{cccccc} 
            1 & 1 & 1 & 0 & 0 & 0\\
            1 & \beta & \beta^2 & 0 & 0 & 0\\
            1 & 0 & 0 & 1 & 1 & 0\\
            1 & 0 & 0 & \beta & \beta^2 & 0\\
            0 & 1 & 0 & 1 & 0 & 1\\
            0 & 1 & 0 & \beta & 0 & \beta^2\\
            0 & 0 & 1 & 0 & 1 & 1\\
            0 & 0 & 1 & 0 & \beta & \beta^2
            \end{array} \right)
 \end{equation}
Based on \textit{Step 2} of the Procedure part in the Algorithm, if the value of $\lceil \frac{s}{r}\rceil$ is $1$, then $W^*$ is equal to $Q^1$.
Therefore, we can construct the parity-check matrix $H$ as shown previously in Equation (\ref{eq:5}) and say the q-ary code $\mathcal{C}$ created with this parity-check matrix is an information $(3, 2, 3)_q$-sequential-locality code.
\end{example}
 

\section{Performance Analysis}

In this section, we aim to analyze our code construction based on the main theorem of this research, focusing on the performance of the algorithm discussed in the previous section.
Before presenting Theorem \ref{the:1}, we need to prove the following lemmas and to simplify understanding the details, let us suppose $t \geq t_i(\delta-1)$ and $\mu=b(\delta-1)$.

\begin{lemma}\label{lem:2}
 Let $\mathcal{C}$ be a q-ary linear $[n, k]_q$ code with a parity check matrix $H$ defined as the output of the algorithm $1$.
 Thus, the following statements hold:
    \begin{enumerate}
       \item Each $i \in [k]$ has $t_i$ disjoint recovery sets, 
       i.e., $L_{j_{\ell}} \cup \{k+j_{\ell}\}\backslash\{i\}$, 
       where $L_{j_{\ell}},\ \ell=1, ..., t_i$, are lines containing $i$.
       
       \item Each $i \in \{k+1, ..., k+\mu \}$ has a recovery set $\mathcal{R} \subseteq [k]$.
       
       \item Each $i \in \{k+1, ..., k+s\}$ has a recovery set $\mathcal{R} \subseteq \{k+1, ..., k+s\} \cup \{k+\mu+1, ..., n\}\backslash \{i\}$.
       
       \item Each $i \in \{k+\mu+1, ..., n\}$ has a recovery set $\mathcal{R} \subseteq \{k+1, ..., k+s\}$.
    \end{enumerate}
\end{lemma}

\begin{proof}
The basic concept of the proof is the same as that of Example \ref{ex:1}.
Claims 1 and 2 are derived from analyzing the first $\mu$ rows of $H$;
Also, claims 3 and 4 are derived from analyzing the last $\lceil\frac{s}{r}\rceil$ rows of $H$. \small $\square$
\end{proof} 

\begin{lemma}\label{lem:3}
   The q-ary linear code $\mathcal{C}$ that has a parity check matrix $H$ in Algorithm $1$ is an $[n=k+\mu +{\lceil{\frac{s}{r}}\rceil}(\delta-1),\ k]_q$-LRC with information $(r, t_i, \delta)_q$-sequential-locality.
\end{lemma}

\begin{proof}
According to the structure of $H$ in (\ref{eq:8}), take the first $k$ columns linked to the $k$ information symbols of $\mathcal{C}$, and the remaining $(b+\lceil\frac{s}{r}\rceil)(\delta-1)$ columns linked to the parity symbols. 
So, $H$ can be partitioned into $b+\lceil\frac{s}{r}\rceil$ row blocks, each consisting of $\delta-1$ rows such that the $j$-th row block, contains rows from the $[(j-1)(\delta-1)+1]$-th to the $j(\delta-1)$-th row, $1\leq j \leq b+\lceil\frac{s}{r}\rceil$. If $1\leq j \leq b$, considering the non-zero columns in the $j$-th row block enables us to derive the q-ary parity-check matrix $\widehat{H}$ associated with the MDS code in \textit{Step 2} of the Inputs in Algorithm $1$. 
Then, puncturing $\mathcal{C}$ through the zero-filled columns of the $j$-th row block yields a q-ary local subcode $\widehat{\mathcal{C}}$ corresponding to $\widehat{H}$ which has length $r+\delta-1$, dimension $r$ and minimum distance $\delta$.
In addition, it contains exactly $\delta-1$ parity symbols that are related to the last $\delta-1$ nonzero columns of $\widehat{H}$.

Moreover, by considering $M^*$ created in \textit{Step 1} of the Procedure section of the Algorithm based on $M$ from \textit{Step 3} of the Inputs, which has uniform column weight $t_i$, it is clear that each of the $i$-th information symbols is included in $t_i$ such q-ary local subcodes for $1\leq i \leq k$, and the supports of these $t_i$ local subcodes only intersect on this coordinate.
Thus, all these information symbols satisfy the conditions of the first part of Definition \ref{def:3}, and the proof is finished. \small $\square$
\end{proof}

\begin{lemma}\label{lem:4}Let $\mathcal{C}$ be the q-ary linear code that has a parity check matrix $H$ defined as the output of the algorithm $1$.
Hence, $\mathcal{C}$ is an $(r, t)_q$-SLRC, where $t \geq 2$ is any positive integer.\end{lemma}

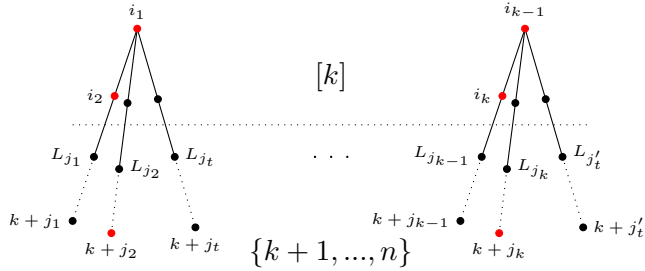
\begin{figure}[ht]
    \centering
    \begin{tikzpicture}[scale=0.85]
    
        \coordinate (A) at (2,3);
        
        \coordinate (N) at (1.65, 1.95);
        \coordinate (E) at (1.33, 1);
        \coordinate (C) at (1,0);
        
        \coordinate (M) at (1.85, 1.84);
        \coordinate (D) at (1.72, 0.81);
        \coordinate (G) at (1.6, -0.19);
        
        \coordinate (L) at (2.32, 1.9);
        \coordinate (F) at (2.58,1);
        \coordinate (B) at (2.9, -0.1);

        \draw (A) -- (E);
        \draw[dotted] (E) -- (C);
        \draw (A) -- (D);
        \draw[dotted] (D) -- (G);
        \draw (A) -- (F);
        \draw[dotted] (F) -- (B);

        \filldraw (A) circle (1.5pt) node[above] {\tiny$i_{1}$}[red];
        
        \filldraw (N) circle (1.5pt) node[left] {\tiny$i_2$}[red];
        \filldraw (E) circle (1.5pt) node[left] {\tiny${L_{j_1}}$};
        \filldraw (C) circle (1.5pt) node[left] {\tiny$k+{j_1}$};
        
        \filldraw (M) circle (1.5pt);
        \filldraw (D) circle (1.5pt) node[right] {\tiny${L_{j_2}}$};
        \filldraw (G) circle (1.5pt) node[below] {\tiny$k+{j_2}$}[red];
        
        \filldraw (L) circle (1.5pt);
        \filldraw (F) circle (1.5pt) node[right] {\tiny$L_{j_{t}}$};
        \filldraw (B) circle (1.5pt) node[below] {\tiny$k+{j_{t}}$};

        \coordinate (M1) at (1, 1.5);
        \coordinate (N1) at (9, 1.5);
        \coordinate (E1) at (5, 2.25);
        \filldraw (E1) node {$[k]$};
        \coordinate (E'1) at (5, -0.5);
        \filldraw (E'1) node {$\{k+1, ..., n\}$};
                 \draw[dotted] (M1) -- (N1);

\coordinate (A1) at (4.75,1);
\filldraw (A1) circle (0.25pt);
\coordinate (B1) at (5,1);
\filldraw (B1) circle (0.25pt);
\coordinate (c1) at (5.25,1);
\filldraw (c1) circle (0.25pt);

        \coordinate (A') at (8,3);
        
        \coordinate (N') at (7.65, 1.95);
        \coordinate (E') at (7.33, 1);
        \coordinate (C') at (7,0);
        
        \coordinate (M') at (7.85, 1.84);
        \coordinate (D') at (7.72, 0.81);
        \coordinate (G') at (7.6, -0.19);
        
        \coordinate (L') at (8.32, 1.9);
        \coordinate (F') at (8.58,1);
        \coordinate (B') at (8.9, -0.1);

        \draw (A') -- (E');
        \draw[dotted] (E') -- (C');
        \draw (A') -- (D');
        \draw[dotted] (D') -- (G');
        \draw (A') -- (F');
        \draw[dotted] (F') -- (B');

        \filldraw (A') circle (1.5pt) node[above] {\tiny$i_{k-1}$}[red];
        
        \filldraw (N') circle (1.5pt) node[left] {\tiny{$i_k$}}[red];
        \filldraw (E') circle (1.5pt) node[left] {\tiny$L_{j_{k-1}}$};
        \filldraw (C') circle (1.5pt) node[left] {\tiny$k+{j_{k-1}}$};
        
        \filldraw (M') circle (1.5pt);
        \filldraw (D') circle (1.5pt) node[right] {\tiny$L_{j_k}$};
        \filldraw (G') circle (1.5pt) node[below] {\tiny$k+{j_k}$}[red];
        
        \filldraw (L') circle (1.5pt);
        \filldraw (F') circle (1.5pt) node[right] {\tiny$L_{j'_{t}}$};
        \filldraw (B') circle (1.5pt) node[right] {\tiny$k+{j'_{t}}$};

    \end{tikzpicture}
    \caption{\centering Points and recovery sets are shown as follows.}
    \label{fig:1}
\end{figure}

\begin{proof}
According to the second part of Definition \ref{def:3}, we need to show that Proposition \ref{pro:1} holds.
It means that for any $\mathcal{I} \subseteq [n]$ of size $|\mathcal{I}| \leq t$, there exists an $i \in \mathcal{I}$ such that $i$ has a recovering set $\mathcal{R}_i \subseteq [n]\backslash \mathcal{I}$.
To simplify the proof, we can analyze the different positions in which $\mathcal{I}$ is placed in $[n]$ and apply Lemma \ref{lem:2}.
Since the proving process is similar to Theorem $33$ in \cite{bib8}, we omit some details.

\begin{description}
    \item[P1:] $\mathcal{I} \cap [k] = \emptyset$. 
    Therefore, we have $\mathcal{I} \cap \{k+1, ..., k+\mu\} \neq \emptyset$, or else, $\mathcal{I}\subseteq\{k+\mu+1, ..., n\}$,
    so, by $(2)$ of Lemma \ref{lem:2}, $i$ has a recovering set $\mathcal{R}_i\subseteq [k]\subseteq [n]\backslash \mathcal{I}$, if not, by $(4)$ of Lemma \ref{lem:2}, $i$ has a recovering set $\mathcal{R}_i \subseteq \{k+1, ..., k+s\}\subseteq [n]\backslash \mathcal{I}$.
   
    \item [P2:] $\mathcal{I} \cap [k] \neq \emptyset$. 
    Take an $i_1 \in \mathcal{I}\cap [k]$. 
    Let 
    \begin{equation}\label{eq:11}
          \mathcal{R}_{\ell}=L_{j_{\ell}}\cup\{k+j_{\ell}\}\backslash\{i_1\},\ \ell=1, ..., t_i,  
    \end{equation}
    such that $L_{j_{1}}, ..., L_{j_{t_i}}$ represent the $t_i$ lines containing $i_{1}$, from $(1)$ of Lemma \ref{lem:2}, $\mathcal{R}_1, ..., \mathcal{R}_{t_i}$ are $t_i$ disjoint recovery sets of $i_1$.
    If $\mathcal{R}_{\ell}\subseteq [n]\backslash \mathcal{I}$ for some $\ell \in [t_i]$, it's done. 
    If not, suppose $\mathcal{I}\cap \mathcal{R}_{\ell} \neq \emptyset$ for all $\ell \in [t_i]$.
    Since all recovery sets are disjoint, so $|\mathcal{I}|\leq t=t_i(\delta-1),\ |\mathcal{I}\cap \mathcal{R}_{\ell}|\leq (\delta-1),\ \ell=1, ..., t_i$, and $\mathcal{I}\subseteq\{i_1\}\cup(\bigcup_{\ell=1}^{t_i}\mathcal{R}_{\ell})$. We have the following cases:
\end{description}

\begin{description}
    \item [$P_{2.1}$]: $\mathcal{I}\cap \mathcal{R}_{\ell}\subseteq [k], \forall \ell \in \{1, ..., t_i\}$. Then $\mathcal{I}\subseteq [k]$, hence from $(1)$ of Lemma \ref{lem:2}, and Lemma \ref{lem:3}, for each $i_1 \neq i_2 \in \mathcal{I}$, $i_2$ has a recovering set $\mathcal{R}_{i_2} \subseteq [n]\backslash[\mathcal{I}]$, such that $\mathcal{R}_{i_1}\cap \mathcal{R}_{i_2} = \{i_1\}.$
    \item [$P_{2.2}$]: $\mathcal{I}\cap \mathcal{R}_{\ell_1}\subseteq [k]$ and $\mathcal{I}\cap \mathcal{R}_{\ell_2}\nsubseteq [k]$ for some $\ell_1, \ell_2 \subseteq \{1, ..., t_i\}$. 
    Take $i_2\in \mathcal{I}\cap \mathcal{R}_1 \subseteq [k]$ and $\mathcal{I}\cap \mathcal{R}_2 \nsubseteq [k]$. 
    Thus, based on (\ref{eq:11}), there is $\mathcal{I}\cap \mathcal{R}_2=\{k+j_2\}$.
    Following (1) of Lemma \ref{lem:2}, $\mathcal{R}'_1, ..., \mathcal{R}'_{t_i}$, are $t_i$ disjoint recovery sets of $i_2$,
    such that $\mathcal{R}'_1=L_{j_1}\cup \{k+j_1\}\backslash \{i_2\}$ and $\mathcal{R}'_{\ell}=L_{j'_1}\cup \{k+j_{\ell}\}\backslash \{i_2\}$, $\ell=2, ..., t_i$.
    It is clear that $L_{j_1}$ is the only line containing $\{i_1, i_2\}$, and hence $\{i_1, i_2, k+j_2\}\cap R'_{\ell}=\emptyset, \ \forall \ell \in \{2, ..., t_i\}$ (see Fig. \ref{fig:1}). So
    from Lemma \ref{lem:3}, there exists an $\ell_{\xi}\in \{2, ..., t_i\}$ such that $|\mathcal{I}\cap \mathcal{R}'_{\ell_{\xi}}|=\emptyset$, and $\mathcal{R}'_{\ell_{\xi}}\subseteq [n]\backslash \mathcal{I}$ is a recovery sets of $i_2$.
    \item [$P_{2.3}$]: $\mathcal{I}\cap \mathcal{R}_{\ell} \nsubseteq [k]$ for all $\ell \in \{1, ..., t_i\}$. 
    Hence $\mathcal{I}\cap \mathcal{R}_{\ell}=\{k+j_{\ell}\}$,
    and from Definition \ref{def:5}, $L_{j_1}, ..., L_{j_{t_i}}$ are all lines that contain $i_1$ which belong to the $s$ parallel classes.
    Suppose for all $\ell \in \{1, ..., t_i\}$, $L_{j_{\ell}} \in \mathcal{L}_{\ell}$,
    therefore $j_1 \leq s$ and $s<j_{\ell}\leq \mu, \ \ell=2, ..., t_i$.
        Considering (3) of Lemma (\ref{lem:2}), $k+{j_1}$ has a recovering set $\mathcal{R} \subseteq \{k+1, ..., k+s\}\cup\{k+\mu +1, ..., n\}\backslash \{j_1\}\subseteq [n]\backslash \mathcal{I}$.
 \end{description}
 Considering the above discussion, for any $\mathcal{I}\subseteq [n]$ of size $|\mathcal{I}|\leq t$, there exists an $i \in \mathcal{I}$ in which $i$ has a recovering set $\mathcal{R} \subseteq [n]\backslash \{\mathcal{I}\}$.
 Therefore, based on Lemma \ref{lem:1}, the proof is completed. \small $\square$

\end{proof}

\begin{theorem}\label{the:1}
The q-ary code $\mathcal{C}$, which has a parity check matrix $H$ in (\ref{eq:8}), is an $(r, t)_q$-SLRC with information $(r, t_i, \delta)_q$-sequential-locality, where $t$ can be any number of erasures and

\begin{align}\label{eq:12}
\frac{k}{n} =\left(1+\lceil\frac{t}{r}\rceil+\lceil\frac{1}{r^2}\rceil(\delta-1)\right)^{-1}.
\end{align}\end{theorem}

\begin{proof}According to what was previously discussed, the first part of the theorem is obvious.
By the construction, $\mathcal{C}$ has block length
       \begin{align*}
         n
         & =k+\mu+\lceil\frac{s}{r}\rceil(\delta-1)\\
         & =k+b(\delta-1)+\lceil\frac{s}{r}\rceil(\delta-1) \\
         & =k(1+\lceil\frac{t_i}{r}\rceil(\delta-1)+\lceil\frac{1}{r^2}\rceil(\delta-1)).
       \end{align*}
Let $t=t_i(\delta-1)$, so the code rate is
       \begin{align*}
        \frac{k}{n}= \left(1+\lceil\frac{t}{r}\rceil+\lceil\frac{1}{r^2}\rceil(\delta-1)\right)^{-1}.
       \end{align*}
\end{proof}


\section{Comparison to other LRCs}

The majority of work in the general literature on LRCs focuses on binary alphabets, parallel recovery, or both.
Moreover, almost all of the research is focused on codes with a small number of erasures (e.g., $t = 2$) \cite{bib20}.
On the other hand, our code structure is more general and proposes a new connection between parallel and sequential recovery, which leads to a construction for both binary and/or non-binary $t$-seq LRCs, for any field size $q$ and any number of erasures $t \geq 2$.
Although the information rate of the code constructed in theorem \ref{the:1} is the lowest for an SLRC with locality $r$, we show that it performs well in the structure through the following comparisons.

\subsection{Code Constructions}
   
\begin{description}
    \item \textit{Graph:} In \cite{bib11}, optimal $t$-seq LRCs are derived from graph construction-based for the case of binary $2$-seq LRCs.
    Similarly, \cite{bib2} extended these results to the $3$-seq LRCs by using hypergraphs.
    In \cite{bib13} and \cite{bib8}, optimal LRCs are built from Moore graphs and resolvable configurations for binary $t$-seq LRCs, respectively.
    However, for the first family, these corresponding graphs only exist for $t \in \{2, 3, 4, 5, 7, 11\}$ when $r \geq 2$, and for the second one when $t$ is odd.
    In contrast, our code family is more general to repair $t$-seq LRCs for any erasure nodes, removes the limitation on locality $r$, and works for any field size $q$.   
\end{description}
\begin{description}
   \item \textit{Code Products:} The authors of \cite{bib8} introduced products of binary $[r, r+1]$-MDS codes, and \cite{bib20} generalized this to $t$-seq LRCs over $q$-ary fields, which can recover any number of erasures by a small locality $r$.
   Compared to them, we establish a new connection between parallel recovery and sequential recovery, which yields $t$-seq LRCs over $q$-ary fields besides linear codes with information $(r, t_i, \delta)$-sequential-locality.
\end{description}
\begin{description}
   \item \textit{Good Polynomials:} In the reference, \cite{bib22} has provided a structure for $2$-seq LRCs that utilizes good polynomials within a small alphabet size of $q$.
   Our structure extends it to any $q$-ary fields and removes the condition on parameter $r$ to be even.
\end{description}
\begin{description}
  \item \textit{Parity-check Matrix:}
   The paper \cite{bib21} linked sequential recovery to linear block codes, showing that a linear block code with girth $2(t+1)$ performs a $t$-seq LRC with locality $r$ if its parity-check matrix has columns weighting at least $2$ and rows weighting at most $r+1$. 
   Our parity-check matrix has accurate parameters. 
   It's a matrix with uniform column weight $t_i(\delta-1)$ and row weight $r+1$, and for the large $t$, with girth \textbf{g}$\geq 4$ is more advantageous.
\end{description}

\subsection{Code Parameters}

To our knowledge, the only study on general $q$-ary sequential locally recoverable codes in the case of binary and/or non-binary for any number of erasure $t\geq 2$ and locality $r$ is presented in \cite{bib20}.
There exists a gap between their upper bounds and what we have achieved. However, the following details help to cover this difference.
Consider the different parameters of the recent SLRCs codes, which are compared in Table \ref{tab:mytable01}.

\begin{description}
    \item[(1)]  Other research has not been able to reach the upper bounds on the total number of recoverable erasures $ \leq q^{t_i}-1$ established in that study.
    However, our construction can achieve more recoverable nodes than the other constructions.
    Let's consider disjoint recovery sets for any code symbol. Comparable to what is proposed in \cite{bib16}, our $t$-seq LRC can recover at least $\delta t_i+1$ up to $t_i \lceil\frac{k}{r}\rceil(\delta-1)+1$ erasures by the same approach, which is significantly more.
    However, our study focuses on the case when the recovery sets are not disjoint and shows that sequential recovery is beneficial over parallel recovery.

    \item[(2)] When \(\delta = 2\), it implies \(t = t_i\), and the bound in equation (\ref{eq:12}) is expressed as \(\left(1 + \left\lceil \frac{t}{r} \right\rceil + \left\lceil \frac{1}{r^2} \right\rceil \right)^{-1}\).
    Therefore, we can derive the information rate similarly to paper \cite{bib8} for any $t \geq 2$, but this time for both binary and non-binary cases under the condition \(q \geq r\).

    \item[(3)] 
    The results obtained show that the total number of erasures that can be recovered in the sequential approach connection with linear block codes is directly related to the girth of the codes.
    For example in \cite{bib21}, if a linear block code has girth $6$, it can be a $2$-seq, or as mentioned in \cite{bib13}, the linear block code must have a girth $\geq 6$ to be a $5$-seq. 
    However, our code construction e.g., with the Tanner graph for $(2, 3)$-regular matrix, has girth $4$ and can recover at least $7$ nodes by sequential approach. 
    This advantage becomes particularly evident in cases involving a large number of erasure nodes.
    
\end{description}

\begin{table}[ht]
\centering
\caption{\centering  Comparison of the various parameters of the SLRCs }
\begin{tabular}{cccccccc}
    \toprule{}
   Ref. &  \begin{tabular}{c}
           \hspace{2.5cm} Parameters \hspace{2.5cm} \\
       \midrule
   \hspace{0.25cm} $t$ \hspace{1.cm} $r$ \hspace{1.25cm} $q$ \hspace{2cm} $n$ \\
      \end{tabular} \hspace{1cm} & Note  \\ 
     \midrule
     \midrule
    \cite{bib8}$_{1}$ & \hspace{0.25cm} any $t$ \hspace{0.4cm} any $r$ \hspace{1cm} $2$ \hspace{1.75cm} $(1+\frac{2}{r})r^m$ \hspace{0.5cm} & $m\in \mathbb{N}^+$\\
     \midrule
      \cite{bib20} & \hspace{0.5cm} any $t$ \hspace{0.6cm} $2$ \hspace{1cm} any $q$ \hspace{1.5cm} $(q+1)^t$ \hspace{1.cm} &  \\
        \midrule
	  \cite{bib22} & \hspace{0.0cm} $2$ \hspace{0.55cm} even $r$ \hspace{0.35cm} $q \thickapprox \frac{(r+1)n}{r+2}$ \hspace{1cm} $(r+2)t_n$ \hspace{0cm} & \\    
      \midrule
        \cite{bib8}$_{2}$ & \hspace{0.05cm} odd $t$ \hspace{0.4cm} any $r$ \hspace{1cm} $2$ \hspace{1.5cm} $k+b+\lceil\frac{s}{r}\rceil$ \hspace{0.25cm} & $s=\lceil\frac{k}{r}\rceil$\\
     \midrule
	  this paper & \hspace{0.5cm} any $t$ \hspace{0.3cm} any $r$ \hspace{0.75cm} $q\geq r$ \hspace{0.75cm} $k+(b+\lceil\frac{s}{r}\rceil)(\delta-1)$ & \\          
    \bottomrule{}
	 \\
	\end{tabular}
 \label{tab:mytable01}
\end{table}

\section{Conclusion}

We developed a general code structure, which is a new connection between parallel and sequential recovery based on the parity-check matrix approach. 
It produced non-binary SLRCs for any field size $q$ and proved that these codes are $(r, t)_q$-SLRCs with information $(r, t_i, \delta)_q$-sequential-locality. That allows the sequential approach to achieve local recovery for at least $t \geq \delta t_i+1$ $(t_i\geq 2)$ erasures each one by $r$ other code symbols.
Besides, our structure obtains better code parameters than previously known constructions with the same property when SLRC is utilized in the block designs.
In the upcoming study, we are interested in exploring more connections between block designs and SLRCs to develop new q-ary SLRCs with sufficient parameters that may improve the information coding rate.

\printbibliography[heading=bibnumbered]

\end{document}